\documentclass[graybox]{svmult}

\usepackage{type1cm}        
\usepackage{makeidx}         
\usepackage{graphicx}        
\usepackage{multicol}        
\usepackage[bottom]{footmisc}
\usepackage{subfig}

\usepackage{newtxtext}       %
\usepackage[varvw]{newtxmath}       

\makeindex             

\begin{document}

\title*{Target Selection for the Redshift-Limited WAVES-Wide with Machine
Learning}
\titlerunning{Target Selection for WAVES wide} 
\author{Gursharanjit Kaur\orcidID{0000-0002-6096-5803}, Maciej Bilicki\orcidID{0000-0002-3910-5809}, Wojciech Hellwing\orcidID{0000-0003-4634-4442}, and the WAVES team }
\authorrunning{G Kaur, M Bilicki, W Hellwing \& WAVES} 
\institute{Gursharanjit Kaur (\email{gursharanjjitk@cft.edu.pl}) 
\and Maciej Bilicki \and  Wojciech A. Hellwing \at Center for Theoretical Physics PAS, al. Lotnik\'{o}w 32/46, 02-668 Warsaw, Poland}

\maketitle

\abstract{The forthcoming Wide Area Vista Extragalactic Survey (WAVES) on the 4-metre Multi-Object Spectroscopic Telescope (4MOST) has a key science goal of probing the halo mass function to lower limits than possible with previous surveys. For that purpose, in its Wide component, galaxies targetted by WAVES  will be flux-limited to $Z<21.1$ mag and will cover the redshift range of $z<0.2$, at a spectroscopic success rate of $\sim95\%$. Meeting this completeness requirement, when the redshift is unknown a priori, is a challenge. We solve this problem with supervised machine learning to predict the probability of a galaxy falling within the WAVES-Wide redshift limit, rather than estimate each object’s 
redshift. This is done by training an XGBoost tree-based classifier to decide if a galaxy should be a target or not. Our photometric data come from 9-band VST+VISTA observations, including KiDS+VIKING surveys. The redshift labels for calibration are derived from an extensive spectroscopic sample overlapping with KiDS and ancillary fields. Our current results indicate that with our approach, we should be able to achieve the completeness of $\sim95\%$, which is the WAVES success criterion.}

\section{Introduction}
\label{sec:1}
The 4-metre Multi-Object Spectroscopic Telescope (4MOST, [1]) is an upcoming fibre-fed facility that will be hosted on VISTA at the Paranal Observatory in Chile. Wide Area Vista Extragalactic Survey (WAVES, [2]), one of the key extragalactic campaigns on 4MOST, is primarily a galaxy evolution survey which aims at measuring the spectroscopic redshifts of about 1.6 million galaxies covering a sky area of $\sim$ 1200 $ \mathrm{deg}^{2}$ . 

The main scientific goal of WAVES is to study the halo mass function at significantly lower galaxy and halo masses in the low-redshift Universe as compared to previous surveys like the Galaxy And Mass Assembly (GAMA, [3]).  In WAVES we plan to do so in two sub-surveys: Wide and Deep. Here we focus on the Wide survey (WW), with the fiducial spectroscopic selection being $Z$-band magnitude (central wavelength 0.88 $\mu m$) upper limit of 21.1,  and redshift limit of 0.2 [2]. 
The success of the survey will be determined by its completeness, with a target value of 95\% or higher. 
The flux-limited selection is easy to achieve, as fluxes are readily available from internal WAVES reprocessing of the KiDS+VIKING imaging data. However, redshift-limited selection for a survey which by design aims at measuring the redshifts themselves is challenging. The general approach for such a selection would be photometric redshift-based selection. 

The photometric redshift can be generally estimated via template-fitting or machine-learning methods. However, irrespective of the approach, the resulting scatter is unlikely to be significantly smaller than $\sigma_z \sim 0.02(1+z)$ for the broad-band photometry we have [4]. At the redshift selection limit of 0.2, this would roughly translate to a completeness of 90\%, which is much lower than what is desired by the survey.

Also, for the sake of WW target selection, we only need to know if the galaxy lies below the redshift of 0.2 or not, rather than estimate the redshift directly. Therefore, we employ a supervised machine-learning classification algorithm, XGBoost, to estimate the probability of the galaxy lying within the redshift limit. 
\section{Data}
\label{sec:2}

\subsection{Input photometry}\label{sec:2.1}
The input imaging is from KiDS [5] $ugri$ and VIKING [6] $ZYJHK_s$ as used to produce the KiDS final Data Release 5 [7].  The original imaging has been reprocessed for WAVES purposes with ProFOUND [8] using an approach similar to that of Bellstedt et al. (2020) [9]. The star-galaxy separation for the input catalogue has been described in Cook et al. (2024) [10]. 
To avoid the issue of data missing at random for the machine learning algorithm, we only select galaxies having detections in all the 9 photometric bands. After applying all the cuts, we are left with $14,060,220$ galaxies in the flux limited selection ($Z<21.1$ mag). 

\subsection{Spectroscopic labels}\label{sec:2.2}
The input photometric catalog was cross-matched with overlapping spectroscopic surveys such as GAMA, SDSS, zCOSMOS, etc. The input spectroscopic surveys and datasets are identical to those detailed in Jalan et al. (2024) [11] and Wright et al. (2024) [7]. We combine them into a single parent catalog following the method outlined in the former paper.

The spectroscopic cross-match contains $572,325$ galaxies at the fiducial $Z$-band magnitude limit of 21.1, and has a median redshift of 0.25. This spectroscopic sample serves as the training and test set for the machine learning classifier. 

\section{Methodology}
\label{sec:3}
As previously mentioned, we address the problem of WW target selection as a classification task. Specifically, we assign class `0' to the desired targets (those below the specified $Z$-magnitude and redshift thresholds) and class `1' to those above the redshift limit. To achieve this, we adopt a supervised machine learning approach, trained on galaxies with true labels derived from the spectroscopic data of Sec.~\ref{sec:2.2}. More specifically, we utilize a tree-based algorithm, which is both computationally efficient and well-suited for binary classification tasks. Additionally, our method enables the estimation of the probability of an object being assigned to a particular class.

For target selection, we employ the Extreme Gradient Boosting (XGBoost) classifier using Python's \texttt{scikit-learn} library [12]. The \texttt{xgboost} library is integrated via the Scikit-Learn API to implement the XGBoost model.

We split the spectroscopic cross-match into training and testing sets. 
We use the XGBoost classifier with default hyperparameters. To check for the efficiency of the classification of Class `0', we used Purity, Completeness, and F1 score (harmonic mean of the two former) as the metrics. 

\section{Results}
\subsection{Feature Importance}
\begin{figure}%
    \centering
    \subfloat[\centering ]{{\includegraphics[scale=0.27]{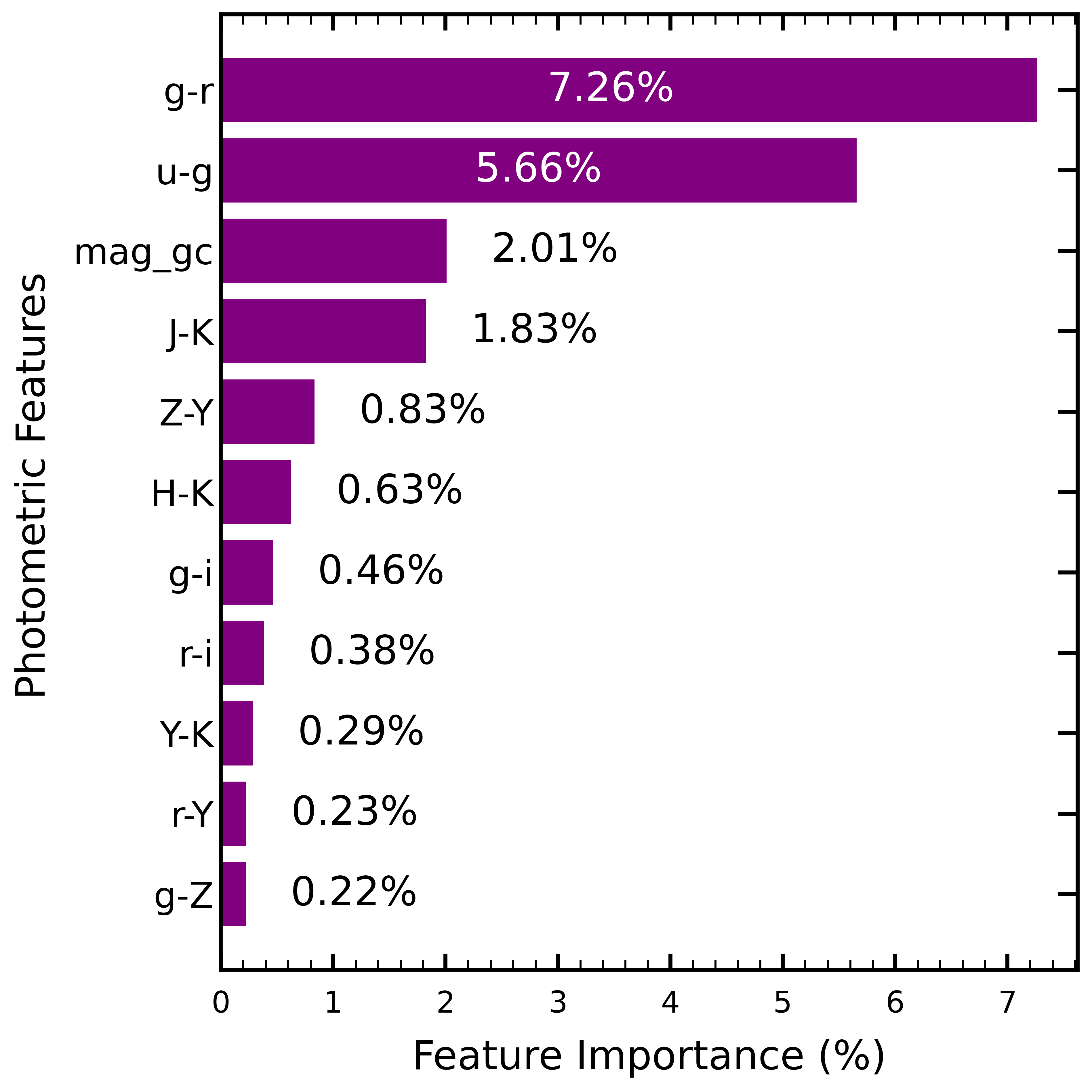} }}%
    \qquad
    \subfloat[\centering ]{{\includegraphics[scale=0.3]{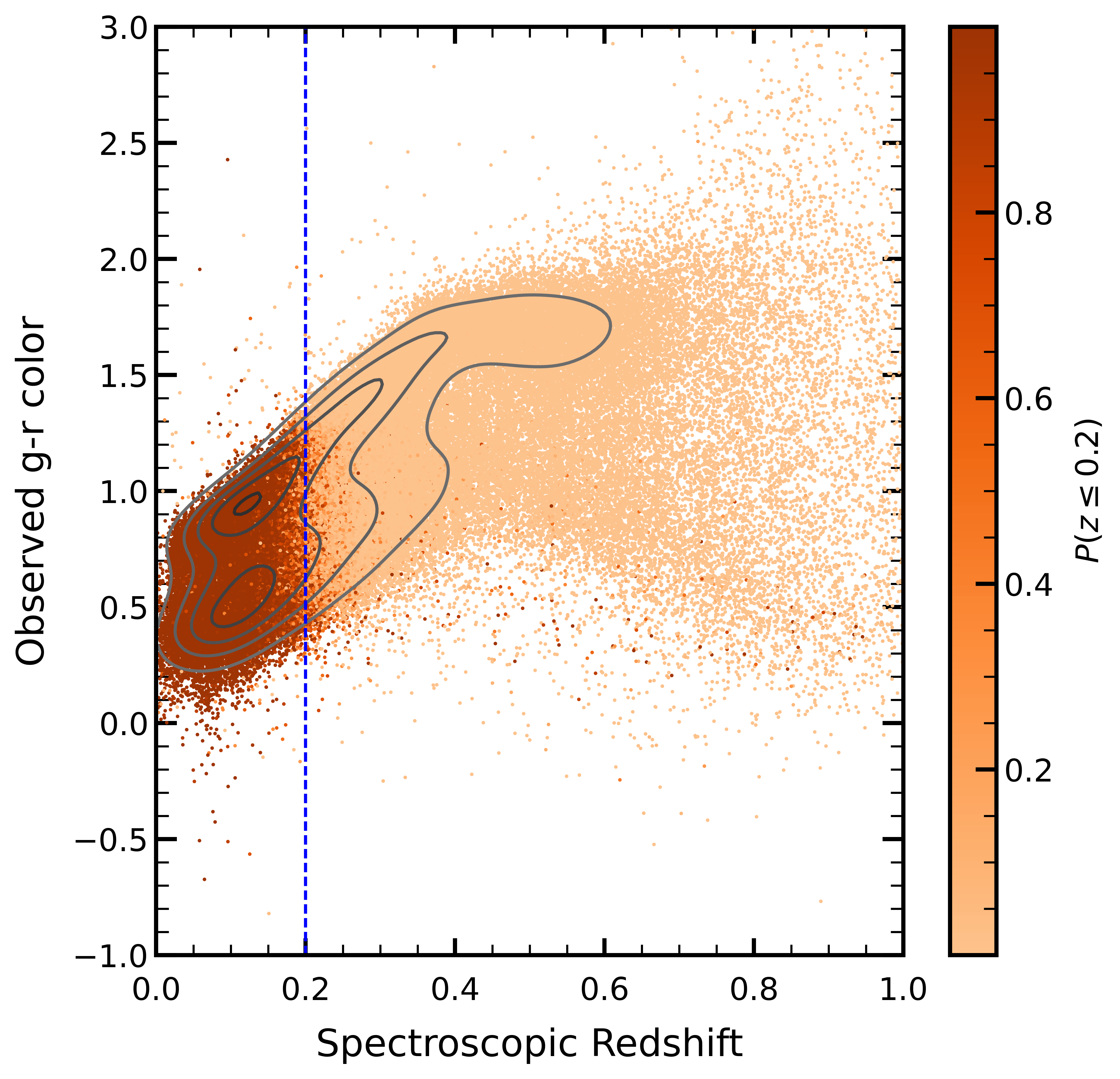} }}%
    \caption{(a) Feature importance for our classifier, computed as relative decrease in the F1 score when the features are shuffled. (b) Distribution of the galaxies from the test set on the true redshift -- observed $g-r$ color plane, color-coded by the probability of being a WAVES-Wide target as assigned by our classifier. The vertical dashed line is the redshift threshold.}
    \label{fig:feature_color}
\end{figure}

Our features are 9-band $ugriZYJHK_s$ magnitudes and all the possible colors. In order to find out the most important features for the classification, we shuffled the values of each feature in the test 30 times and measured the average decrease in the F1 score. The feature importance scores for the 11 most important features are shown in Fig.~\ref{fig:feature_color}(a). It is clear from the figure that the $g$ magnitude \text{mag\_gc} and the associated $g-r$ and $u-g$ colors are the most important. This is likely due to the fact that for the WW limiting redshift of 0.2, the 4000 \AA\ break passes through the VST $g$-band filter.

\subsection{Classification metrics}
The purity, completeness, and F1 score for the classification are all $\sim 94.7\%$. All the results described are for the probability threshold of classification being 0.5, i.e. we consider a galaxy as a target if its $P(z \leq 0.2) \geq 0.5$. The completeness (purity) of our target assignment can be improved by lowering (enlarging) the classification probability threshold. 

Fig.~\ref{fig:feature_color}(b) shows the $g-r$ color vs. spectroscopic redshift of the test set galaxies color-coded by the classification probabilities assigned by the XGB classifier. The misclassified galaxies are light orange to the left of the $z=0.2$ vertical line, and brown to the right of this threshold. Most of the 5.3\% missed out (false negative) galaxies are lying close to the redshift limit.  

We applied the XGB model trained on the full spectroscopic catalog from Sec.~\ref{sec:2.2} to the photometric catalog of Sec.~\ref{sec:2.1} which is our inference/target set. About 2.9 million galaxies from $\sim14$ million total sources in the photometric catalog are assigned as Class `0' (the target galaxies) and have probabilities, $P (z \leq 0.2) \geq 0.5$. This number can be lowered if we increase the probability threshold.

\section*{Acknowledgments}
This work was supported by the Polish National Science Center through grants number 2020/38/E/ST9/00395 (GK \& MB) and 2020/39/B/ST9/03494 (MB \& WAH).

\end{document}